\documentclass[12pt]{iopart}
\usepackage{iopams}
\usepackage{setstack}
\usepackage{graphicx}
\newcommand{\fslash}[1]{\mbox{$\!\not\!#1$}}

\begin{document}
                                                                                
\baselineskip 4 ex

\title{Nucleons, Nuclear Matter and Quark Matter: A unified NJL approach.}
                                                                                
\author{S. Lawley$^{1,2}$, W. Bentz$^3$ and A. W. Thomas$^2$}

\address{$^1$ Special Research Centre
        for the Subatomic Structure of Matter, \\
        University of Adelaide, Adelaide SA 5005, Australia}
\address{$^2$ Jefferson Lab, 12000 Jefferson Avenue,
        Newport News, VA 23606, U.S.A.}
\address{$^3$ Department of Physics, School of Science,
        Tokai University \\
        Hiratsuka-shi, Kanagawa 259-1292, Japan}
\ead{slawley@jlab.org}
                                                                                
\date{ }

\begin{abstract}
 We use an effective quark model to describe both hadronic matter and
deconfined quark matter.
By calculating the equations of state and the corresponding neutron star
properties, we show that the internal properties of the nucleon have
important implications for the properties of these systems.
\vspace{1pc}
                                                                                
\noindent
{\footnotesize PACS numbers: 12.39.Fe; 12.39.Ki; 21.65.+f; 97.60.Jd }
                                                                                
\end{abstract}

\maketitle

\section{Introduction}
\setcounter{equation}{0}
                                                                                
Several recent studies have shown the advantages of using composite
nucleons in finite density calculations \cite{BT01}, \cite{QMC88},
\cite{RP05}.
For example, in some models the saturation mechanism of the system can be
related to how the structure of the nucleon changes with density
\cite{TGLY04}.
While saturating nuclear systems can also be readily obtained using
point-like nucleons \cite{QHD},
if one wishes to explore the transition to quark matter it is desirable to
start with a description of the nucleon on the quark level.
Although the nucleon is a complex object,
some simplification arises by treating it as an interacting
quark-diquark state in the Faddeev approach.
It turns out that this
picture of the nucleon has considerable appeal, both theoretical and
observational\footnote{For a discussion on the evidence in favor of
diquarks see Ref. \cite{Wilczek}.}.
For example, it nicely reproduces the light baryon spectrum \cite{Elena},
while a calculation without diquark correlations predicts an abundance
of missing resonances \cite{Isgur}.
Indeed, the widespread consensus that quark pairing is favorable in high
density quark matter \cite{List}, where perturbative QCD has some predictive power,
lends support to the idea that diquark correlations
within the nucleon are important too \cite{Wilczek}.

In this work we consider the possibility of a connection between the
diquark correlations within the nucleon and the pairing of quarks anticipated
in color superconducting quark matter.
We work with the assumption of two flavors.
At zero temperature and normal nuclear matter density strange quarks
certainly do not feature in the dynamics of the system, so this is a
reasonable starting point.
At higher densities strangeness could play a role through the formation
of hyperons, kaon condensates and/or strange quark matter, depending on
which of these phase transitions is favored and in what sequence they
appear with respect to density \cite{Bandyopadhyay05,Alford05,Weber05,Ping05}
\footnote{The possibility of 3-flavor quark stars was first investigated in 1970 by Itoh \cite{Itoh70}. }.
Actually, there are numerous possibilities for quark pairing and
condensation in
high density matter, especially if strangeness is introduced.
Although these
ideas can all be investigated within the framework we are using,
we wish to concentrate here on the question of whether a unified
description of diquark interactions can be achieved for both
the nuclear matter (NM) and quark matter (QM) phases. For this purpose,
we consider the simplest possibility that two flavor hadronic matter goes
directly to two flavor
QM, using the Nambu-Jona-Lasinio (NJL) model to describe both phases.
In NJL type models, it has been shown that the
mass of the strange quark
tends to inhibit the formation of strange QM in the density
region relevant to compact stars
\cite{Buballa05,David05,Ruster05,Buballa03}.
The possibility of pseudoscalar condensation
is considered in Refs.\cite{He05,Barducci05, Ebert05, Maruyama05},
and the competition between the pseudoscalar condensates and the two-flavor
superconducting phase has also been examined \cite{Warringa05}, although
this remains to be applied to the charge neutral equation of state.
%

The NJL model was originally developed to describe interactions between
structureless nucleons \cite{NJL}.  It has been shown that this approach
does not lead to nuclear stability
\cite{buballa96,koch86,daProvidencia82}.
The idea of the NJL model in its present form is to describe QCD at low
energies by assuming point-like interactions between quarks.
This approximation is justified by the fact that gluon degrees of freedom
should be of relatively minor importance at low energies.  Therefore it is
sufficient to construct a model where the gluon interactions are ``frozen
in,'' meaning they are only present implicitly through the couplings of
the model.
The general form of the NJL Lagrangian density is given by,
\begin{eqnarray}
{\cal L}_{NJL} = \bar{\psi}(i\fslash\partial-m)\psi + \sum_{\alpha}
G_{\alpha} (\bar{\psi} \Gamma_{\alpha} \psi )^{2},
\end{eqnarray}
where $\bar{\psi}$ and $\psi$ are the quark fields, $m$ is the current
quark mass and  $G_{\alpha}$ are the coupling constants associated with
the various interaction channels.
The interaction Lagrangian density can be expressed in a variety of equivalent
forms using Fierz transformations \cite{Ishii95}.  Here we decompose it
into $\bar{q} q $ and $q q$ interaction terms as follows:
\begin{eqnarray}
 {\cal L}^{\rm int}_{\rm NJL} &=&
G_{\pi}(\bar{\psi} \psi )^{2}- G_{\pi}(\bar{\psi} \gamma_{5}\bold{\tau}
\psi )^{2} \\
&-& G_{\omega}(\bar{\psi} \gamma^{\mu} \psi )^{2}
 - G_{\rho}(\bar{\psi} \gamma^{\mu}\bold{\tau} \psi )^{2}\\
&+& G_{s}(\bar{\psi} \gamma_{5} C \tau_2 \beta^A \bar{\psi}^T)(\psi^T
C^{-1} \gamma_5 \tau_2 \beta^A \psi) \\
&+& G_{a}(\bar{\psi} \gamma_{\mu} C \bold{\tau} \tau_2 \beta^A
\bar{\psi}^T)(\psi^T C^{-1} \gamma_\mu \tau_2 \bold{\tau} \beta^A \psi)
\label{lag}
\end{eqnarray}
In order of appearance the above terms correspond to the
scalar, pseudoscalar, vector-isoscalar and vector-isovector $\overline{q} q$
interactions,
and the last two terms correspond to the scalar and the axial
vector diquark channels.
The color matrices are given by
$\beta^{A} = \sqrt{3/2} \, \lambda^{A}$ $(A=2,5,7)$,
indicating that these are attractive, color anti-symmetric diquark channels.
In the NM
phase the diquark interactions will lead to color singlet nucleons, and
in the QM phase to color superconducting pairs.
Indeed, QCD supports the idea that there is a strong attraction in the color
anti-symmetric flavor singlet scalar channel, leading to the formation
of condensed scalar diquarks (the so called 2SC phase) \cite{Wilczek}.
In our model these condensed pairs should arise from the same interaction
(1.4) as the scalar diquarks in the nucleon constructed in the Faddeev
approach.
\footnote{Note that the $SU(3)_{c}$ color symmetry is broken in the
2SC phase - the
pairing is between just two colors and the third color remains unpaired.
We do not assume axial vector diquark condensation in QM, as
this would break further symmetries.}
                                                                                
Because the NJL model is non-renormalizable, the model is only fully
specified after the choice of regularization scheme \cite{Klevansky92}.
In our work quark confinement is ensured mathematically through the
introduction of a finite infra-red cut-off ($\Lambda_{\rm IR}$) in
the proper time regularization scheme \cite{Ebert05,Hellstern98}.
In this method, the unphysical quark decay
thresholds are eliminated \cite{BT01}, as also
found in the Dyson-Schwinger approach \cite{Hellstern98}.
Of course, in the deconfined phase the infra-red cut-off will be set to
zero \cite{Bentz03}.

\section{Nucleons and nuclear matter}
\setcounter{equation}{0}
                                                                                
To describe NM we first consider the nucleon and its internal degrees of
freedom.  Incorporating the quark substructure of the nucleon through the
Faddeev approach will allow us to examine how the nucleon properties change
with density.
The homogeneous Faddeev equation for the vertex function $\Gamma_N$
in the nucleon channel $(J,T)=(1/2,1/2)$ has the form
$\Gamma_N = Z \, \Pi_N\, \Gamma_N$, where $Z$ is the
quark exchange kernel and $\Pi_N$ the product of the quark and
diquark propagators \cite{Ishii95}.
For our present, finite density calculations we restrict ourselves to the
static approximation \cite{Buck}, where the momentum dependence of the
quark exchange kernel $Z$ is neglected and a static parameter, c is instead introduced to reproduce the main features of the exact Faddeev calculation \cite{BT01}.  Then $\Pi_N$ becomes effectively the
quark-diquark bubble graph given by
\begin{eqnarray}
 \Pi^{a b}_{N}(p)& =& \int \frac{d^{4}k}{(2 \pi)^{4}} \tau^{a b} (p - k)
S(k) \\
& =&  \int \frac{d^{4}k}{(2\pi)^{4}} \left( { \tau_{s}(p - k)\atop
0}{0\atop \tau_{a}^{\mu\nu} (p - k)} \right) S(k),
\end{eqnarray}
where $\tau_{s}$ and $\tau^{\mu\nu}_{a}$ refer to the scalar and axial
vector components of the diquark t-matrices respectively
\footnote{It is possible to describe the nucleon with just the scalar diquark
channel and this has been considered elsewhere \cite{LBT05}.  As in Ref.~\cite{Cloet}, we use the approximate 'constant + pole' forms of the diquark t-matrices.  }
, and $S(k)$ is the
constituent quark propagator. The nucleon mass follows from the
requirement that
the Faddeev kernel $K \equiv Z \, \Pi_N$ has eigenvalue 1.
In a similar way the mass of the Delta resonance can be calculated
from the pole position in the $(J,T)=(3/2,3/2)$ channel, where only
the axial vector diquark contributes.

Here we wish to use a more accurate description of the nucleon and delta
masses, including
also the pion cloud contributions \cite{TW}. This will enable us to consider
a quantitative
comparison of the scalar pairing found in the confined and deconfined
phases.
We take into consideration the
self-energy contributions corresponding to the $N \to N\pi \to N$ and $N
\to \Delta\pi \to N$ processes for the nucleon and the $\Delta \to
\Delta\pi \to \Delta$ and $ \Delta \to N\pi \to \Delta$ processes for the $\Delta$ \cite{TW}.  The corresponding self-energies are,
\begin{eqnarray}
   \Sigma_{N} = \sigma_{NN}^{\pi} + \sigma_{N\Delta}^{\pi}
\end{eqnarray}
\begin{equation}
   \Sigma_{\Delta} = \sigma_{\Delta\Delta}^{\pi} + \sigma_{\Delta
N}^{\pi},
\end{equation}
where
\begin{eqnarray}
\sigma_{B B'}^{\pi} = \frac{-3
\hspace{1.5mm}g_{A}^{2}}{16\pi^{2}f_{\pi}^{2}} c_{B B'}\int_{0}^{\infty}dk
\frac{k^{4}u^{2}(k)}{\omega(k)[\omega_{B B'} + \omega(k)]}
\label{self}
\end{eqnarray}
where $\omega_{B B'} = (M_{B} - M_{B'})$ is the physical baryon mass
splitting (e.g. $\omega_{N \Delta} = 1232 - 939$ MeV), and $\omega(k) =
\sqrt{k^{2} + m_{\pi}^{2}}$ is the intermediate pion energy. For the
$\pi B B'$ vertex we assume the phenomenological dipole form
\footnote{The dipole cut-off $\Lambda$ should not be
confused with $\Lambda_{IR}$ and $\Lambda_{UV}$ which are the cut-off
parameters on the quark level in this model.} $u(k) =
\Lambda^{4}/(\Lambda^{2}+k^{2})^{2}$.
The coefficients $c_{B B'}$ come from the standard SU(6) couplings (i.e.
$c_{NN} = 1$, $c_{N \Delta} = 32/25$ \cite{Ross}). The baryon masses
including the pion loop contributions are then given by
$M_B = M_B^{(0)} + \Sigma_B$, where $M_B^{(0)}$ are the ``bare'' masses
which follow from the quark-diquark equation.
                                                                                
The size of the self-energy of the nucleon, $\Sigma_{N}$, is not precisely
known.
Calculations indicate it could be up to -400 MeV \cite{Ross,CBM,Pearce,Hecht}.
In the present work $\Sigma_{N}$ is varied by changing the dipole cut-off,
$\Lambda$, within physically acceptable limits,
in order to investigate how the pion cloud of the nucleon
influences the equation of state of the system.

The form of the effective potential in the mean field approximation
has been derived for symmetric NM in Ref.\cite{Bentz03}, starting from the
quark Lagrangian, Eq.(\ref{lag}), and using the hadronization method.
For the present calculations the effective potential is extended to the isospin
asymmetric case, because for neutron star matter
charge neutrality and chemical equilibrium typically lead to an abundance
of neutrons over protons.
The effective potential can be written as follows:
\begin{eqnarray}
V^{\rm NM} = V_{\rm vac} +  V_{N} - \frac{\omega_0^2}{4 G_{\omega}}
- \frac{\rho_0^2}{4 G_{\rho}} - \frac{\mu_e^4}{12 \pi^2}\,, \label{vnm}
\end{eqnarray}
where
\begin{eqnarray}
V_{\rm vac} &=& 12i \int \frac{{\rm d}^4 k}{(2\pi)^4}\,{\rm ln} \,
\frac{k^2-M^2}{k^2-M_0^2}
+ \frac{(M-m)^2}{4 G_{\pi}}
- \frac{(M_0-m)^2}{4 G_{\pi}}
\label{vac}
\end{eqnarray}
is the vacuum term.
The Fermi motion of the nucleons moving in the scalar and vector mean
fields gives rise to the term
\begin{equation}
\hspace{-12mm}
V_N = -2 \sum_{\alpha={\rm p,n}} \int \frac{{\rm d}^3 k}{(2\pi)^3}\,
\Theta(k_{F_{\alpha}}-k) 
\left( \sqrt{k_{F_{\alpha}}^{2}+M_{N}^{2}(M)} - \sqrt{k^{2}+M_{N}^{2}(M)} \right)
\label{vn}
\end{equation}
where $M_N(M)$ is the nucleon mass in-medium, which is the sum of the
bare mass and the pion cloud contribution.
The relations between the chemical potentials, which are the variables
of $V^{\rm NM}$, and the Fermi momenta appearing in Eq.(\ref{vn}) are given by
\begin{equation}
\mu_{n} = \sqrt{M_{N}^2 + k_{F_{n}}^2} + 3 \omega_0 - \rho_0
\label{mu_n}
\end{equation}
%
\begin{equation}
 \mu_{p}=  \sqrt{M_{N}^2 + k_{F_{p}}^2} + 3 \omega_0  + \rho_0
\label{mu_p}
\end{equation}
%
The constituent quark
mass, $M$, and the mean vector fields ($\omega_0$ and $\rho_0$) in NM
are determined
by minimizing the effective potential for fixed chemical potentials
\footnote{In the actual calculation it is easier to minimize the
energy density ${\cal_{E}} = V +
\sum \mu_{\alpha}\rho_{\alpha}$ for fixed densities,
after eliminating the mean vector fields as $\omega_0 = 6 G_{\omega}
(\rho_p + \rho_n)$, $\rho_0 = 2 G_{\rho} (\rho_p - \rho_n)$.}.
The chemical potential of the (massless) electron in the last term of
(\ref{vnm}) is fixed by the requirement of beta equilibrium as
$\mu_e = \mu_n - \mu_p$.

For discussion of the phase structure of this model in Sect.5, we introduce
the chemical potentials associated with baryon number and isospin:
\begin{equation}
\mu_{B} = \frac{1}{2}(\mu_{p} + \mu_{n}); \hspace{10mm}
\mu_{I} = \frac{1}{2}(\mu_{p} - \mu_{n}).
\label{chems}
\end{equation}
The parameters of the model are determined as follows.
We choose $\Lambda_{\rm IR} = 285$ MeV to be of the order of
$\Lambda_{QCD}$.
We calculate
$\Lambda_{\rm UV}$, $m$ and $G_{\pi}$ so as to reproduce
$f_{\pi}=93$ MeV (through the matrix element for pion decay),
$m_{\pi}=140$ MeV (through the Bethe Salpeter equation for the pion) and
constituent quark mass at zero density,
$M_0=400$ MeV (via the gap equation).
The resulting equation of state is not very sensitive to the initial choices
of $\Lambda_{\rm IR}$ and $M_0$.
The parameter $G_{\omega}$ is fixed to give the empirical
binding energy per nucleon of symmetric NM ($E_{B} = 17$ MeV), and the
parameter
$G_{\rho}$ is adjusted to the symmetry energy ($a_{4} = 32$ MeV at $\rho_0
= 0.17 $fm$^{-3}$).
The coupling in the axial vector diquark channel, $G_{a}$, is fixed by
the mass of
the Delta, since in this case the scalar diquark channel does not
contribute. Then the coupling in the scalar diquark channel, $G_{s}$,
is in turn determined by the nucleon mass.
For the static parameter in the quark exchange kernel we use $c = 450$ MeV.
This calculation of baryon masses is carried out for several initial choices
of the dipole cut-off $\Lambda$, which
controls the magnitude of the pion loop contributions to the nucleon
mass, $\Sigma_N$. We will show the results for three different values
of $\Lambda$, leading to $\Sigma_N = -200$ MeV, $-300$ MeV and
$-400$ MeV. The resulting parameter sets are shown in Table 1.
(The first set corresponds to the case where the nucleon and
delta masses are reproduced without pion cloud contributions.)
\begin{table}[hb]
\begin{center}
\begin{tabular}{|c|c|c|c|c|c|c|}
\hline
$\Lambda$ & $\Sigma_{N}$ & $\Sigma_{\Delta}$ & $r_{a}$ & $r_{s}$  &
$r_{\omega}$& $r_{\rho}$ \\ \hline
-       &  0      &  0     & 0.27  & 0.30 & 0.48 & 0.70 \\
710     &  -200   &  -183  & 0.21  & 0.27 & 0.30 & 0.74 \\
803     &  -300   &  -266  & 0.18  & 0.24 & 0.24 & 0.76 \\
877     &  -400   &  -346  & 0.15  & 0.22 & 0.19 & 0.77 \\  \hline
\end{tabular}
\end{center}
\caption{Parameters corresponding to different choices of $\Sigma_{N}$.
We define $r_{\alpha} = G_{\alpha}/G_{\pi}$ ($\alpha=a,s,\omega,\rho$). $\Lambda$, $\Sigma_{N}$ and $\Sigma_{\Delta}$ are in MeV. }
\end{table}
                                                                                
The function $M_N(M)$, which is needed in (\ref{vn}) to minimize the
effective potential for finite density, is calculated by assuming
that the ratio $g_A/f_{\pi}$ in (\ref{self}) is independent of density.
This is supported by a recent analysis of the pion-nucleus optical
potential \cite{Suzuki02}, which has shown that the pion decay
constant in nuclear
matter is reduced by $20\%$, which is the same as the quenching
of $g_A$ derived from Gamow-Teller matrix elements \cite{GT}. The pion mass
in the medium is constrained by chiral symmetry which leads to
the relation\footnote{For the derivation see Ref.\cite{BT01}.  The pion mass in
this relation is defined at zero momentum.}
$m_{\pi}(\rho)^2 = m_{\pi}^2 \cdot M_0/M$. This small enhancement
of the pion mass in the medium is taken into account in the
calculation but its effect is not very important.

\begin{figure*}
\vspace{7mm}
  \begin{center}
    \includegraphics[width=2.8in,height=2.6in]{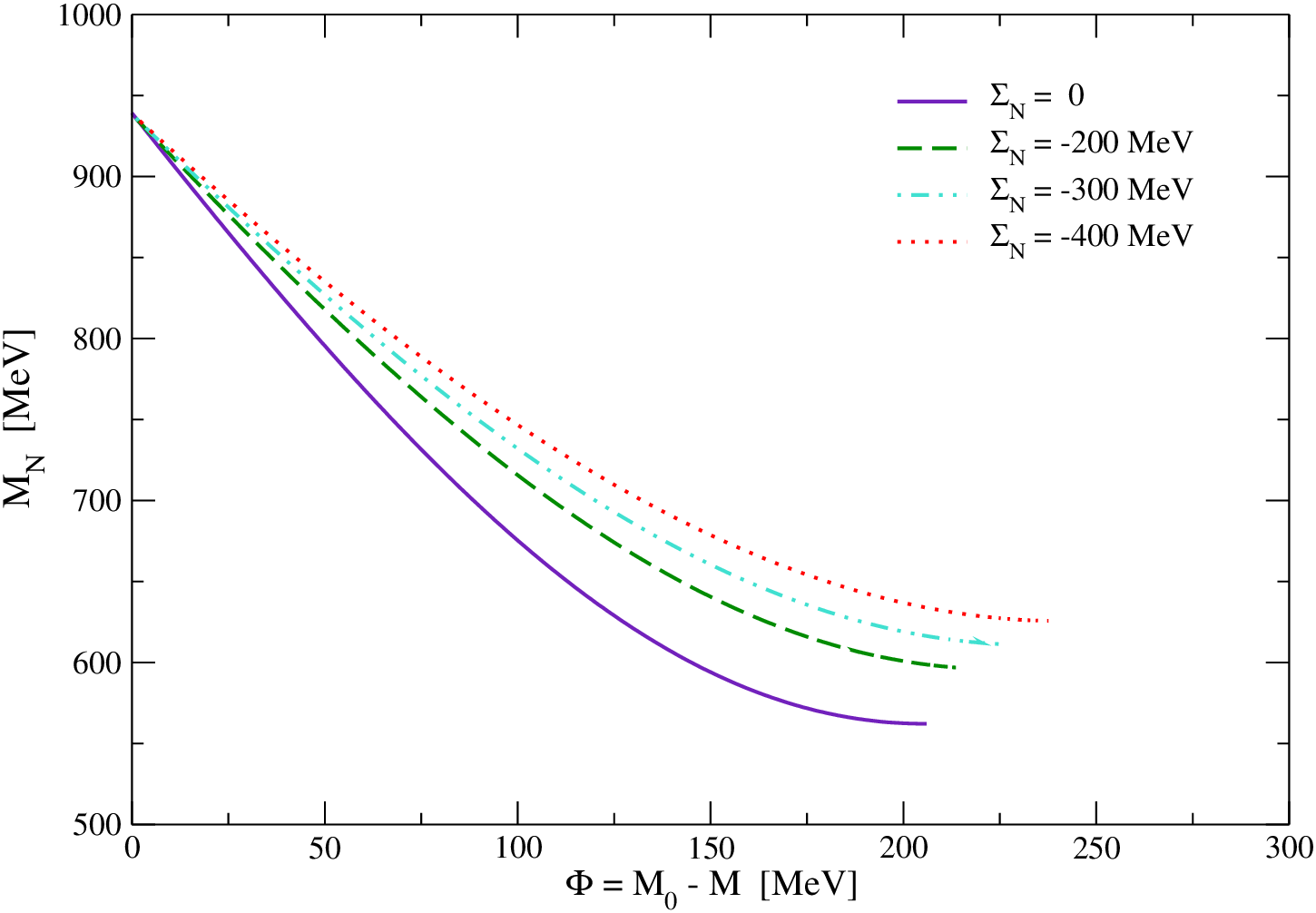}
    \caption{The mass of the nucleon as a function of the scalar potential ($\Phi = M_0 - M$). }
    \label{Polarizability}
  \end{center}
\vspace{2mm}
\end{figure*}
\begin{figure*}
\vspace{10mm}
  \begin{center}
    \includegraphics[width=2.8in,height=2.6in]{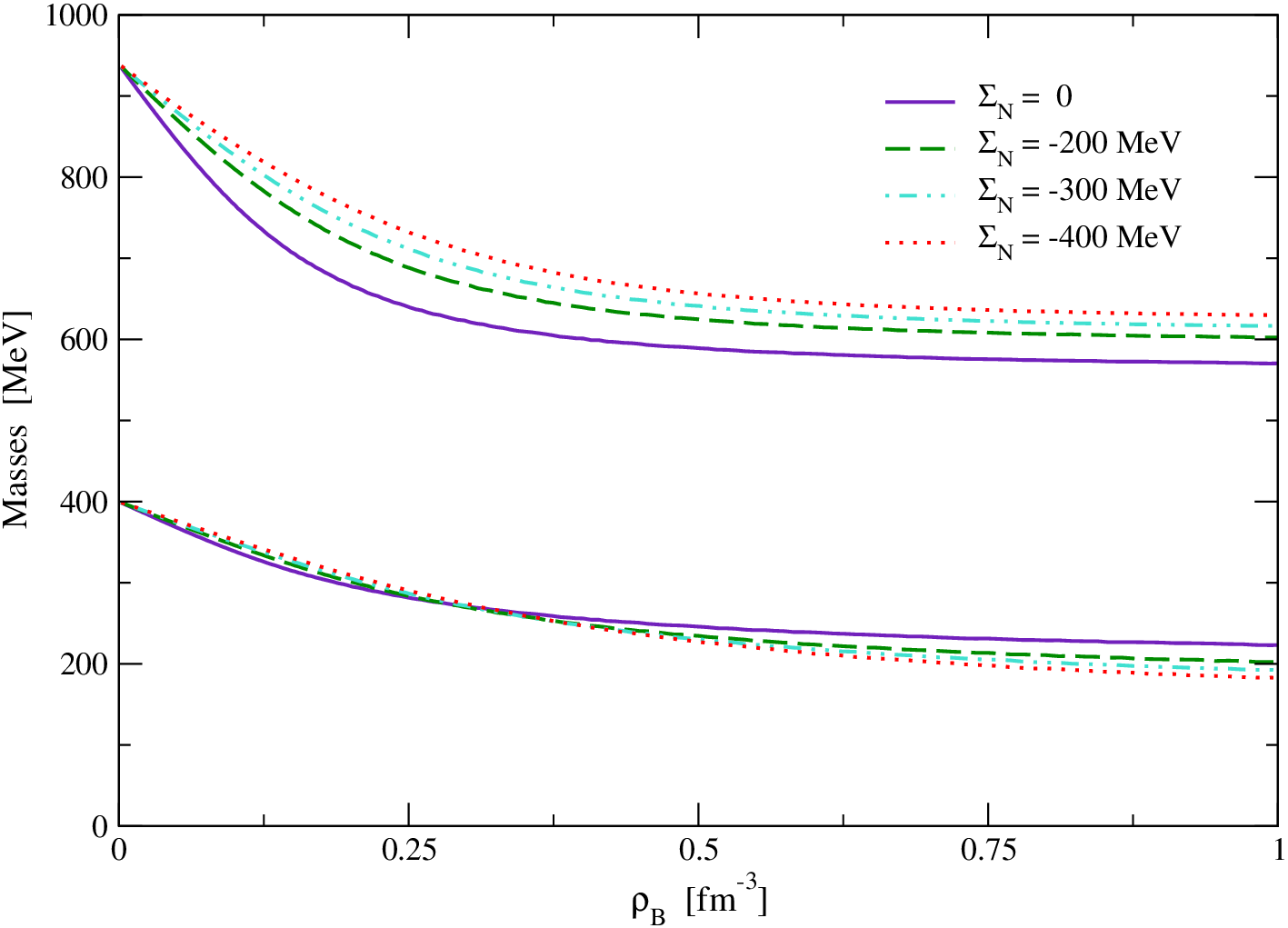}
    \caption{The effective masses of quarks (lower lines) and nucleons (upper lines) in the hadronic phase for several choices of the nucleon self-energy.}
    \label{EffMass}
  \end{center}
\vspace{4mm}
\end{figure*}

                                                                                
The relationship between the nucleon mass and the constituent quark mass
for the four cases of Table 1 is shown in Fig.\ref{Polarizability}.
Note that in the region $M>200$ MeV, which is most relevant for
normal densities, both the slopes and the curvatures of the lines decrease
in magnitude as the attraction from the pion loop increases.
Thus the inclusion of pion loop effects on the nucleon mass leads to
a reduced effective $\sigma\,NN$ coupling and to a reduced
scalar polarizability in the medium \cite{BT01}.  
The self consistently calculated
quark and nucleon masses are shown as functions of the density in
Fig.\ref{EffMass}.

Concerning the pion exchange effects on the NM equation of state, we note that in addition to the term which can be incorporated into the nucleon mass there is also the familiar ``Fock term'' which originates from the Pauli principle.  However this contribution is quite small when the short range correlations between nucleons in the spin-isospin channel are included \cite{EricsonWeise}, and will be neglected here for simplicity.

\section{Quark matter}
\setcounter{equation}{0}
                                                                                
In the most simple mean field description of quark matter, each
energy level for the quarks
is filled up to the Fermi energy. At the Fermi surface, only a small
attraction between quarks leads to the formation of Cooper pairs, analagous
to the phenomenon of BCS pairing of electrons in a superconductor.  In QCD
we anticipate that color superconductivity will arise through the $\bar{3}_{c}$
channel, reducing the energy of the system through the condensation of color
anti-symmetric pairs.
The effective potential for QM thus has the form
\begin{eqnarray}
V^{\rm QM} = V_{\rm vac} +  V_{Q} + V_{\Delta} -
\frac{\mu_e^4}{12 \pi^2} 
\label{vqm}
\end{eqnarray}
where the vacuum part, $V_{\rm vac}$, is the same as for NM (Eq.
(\ref{vac})) except that the infra-red cut-off is zero in QM.  
Furthermore,
\begin{equation}
V_Q = -6 \sum_{\alpha={\rm u,d}} \int \frac{{\rm d}^3 k}{(2\pi)^3}\,
\Theta(\mu_{\alpha} - E_Q(k))\left(\mu_{\alpha} - E_Q(k)\right)
\label{vq}
\end{equation}
describes the Fermi motion of quarks with chemical potentials
$\mu_u$ and $\mu_d$, and $E_Q(k)=\sqrt{M^2+k^2}$. The term (\ref{vq}) is
analagous to $V_{N}$ in NM, except that the quark mass, $M$, corresponds
directly to the scalar field in the system, i.e., there is no scalar
polarizability of the quarks.
The term $V_{\Delta}$
describes the effect of the pairing gap and is given by
\begin{equation}
\hspace{-17mm}
V_{\Delta} = 2i \int \frac{{\rm d}^4 k}{(2\pi)^4}
\sum_{\alpha=+,-} \Big[ {\rm ln} \, \frac{k_0^2 -
(\epsilon_{\alpha}+\mu_I)^2}
{k_0^2 - (E_{\alpha}+\mu_I)^2} \\
+ {\rm ln} \, \frac{k_0^2 - (\epsilon_{\alpha}-\mu_I)^2}
{k_0^2 - (E_{\alpha}-\mu_I)^2} \Big] + \frac{\Delta^2}{6 G_s}\,,
\label{delta}
\end{equation}
where $\epsilon_{\pm}(k) = \sqrt{(E_{Q}(k) \pm \mu_B/3)^2 + \Delta^2}$,
$E_{\pm} = |E_{Q}(k) \pm \mu_B/3|$, and we introduced the chemical
potentials for baryon number and isospin
\footnote{We mention that in principle one
needs a further chemical potential for color neutrality ($\mu_8$) in QM.
However, for the 2-flavor case $\mu_8$ turns out to be very small
\cite{Buballa05,Grig04,ABG}.}
\begin{equation}
\mu_{B} = \frac{3}{2}(\mu_{u} + \mu_{d});\hspace{10mm}
\mu_{I} = \frac{1}{2}(\mu_{u} - \mu_{d}),
\end{equation}
which corresponds to (\ref{chems}) in the NM phase. The electron
chemical potential is determined from beta equilibrium as
$\mu_e = \mu_d - \mu_u = -2 \mu_I$.
                                                                                
The gap $\Delta$ and the quark mass $M$ are determined by minimizing the
effective potential for fixed chemical potentials. Our results, discussed
below, show that $M$ is quite small in the QM phase, i.e., we have
almost current quarks.
                                                                                
Note that the vector-type interactions are set to zero in QM, even though
in the description of NM the vector mean fields are clearly important.
This assumption, which has been made implicitly in almost all
investigations of high density quark matter, is supported by
the discussions of vector meson
poles in Ref.\cite{Bentz03}. It is also supported by recent arguments related
to the EMC effect \cite{Will}, which show that in the high energy region,
where one has essentially current quarks (as in the present high density
case), the mean vector field must indeed be set to zero.

In the following discussions, we distinguish the normal quark matter
(NQM) phase, which is characterized by $\Delta=0$, from the color superconducting (SQM) phase ($\Delta>0$).
Note that the value of $G_{s} = r_{s} G_{\pi}$ controls the outcome of
this minimization through the last term in (\ref{delta}).  In these calculations we focus on the case where the
pairing strength ($r_{s}$) in QM takes the same value as in NM, namely
the value required to obtain the correct nucleon mass after the pion cloud
contribution is taken into account.
%
                                                                                 \begin{figure}
  \begin{center}
    \vspace{17.5mm}
    \includegraphics[width=4.6in,height=4.6in]{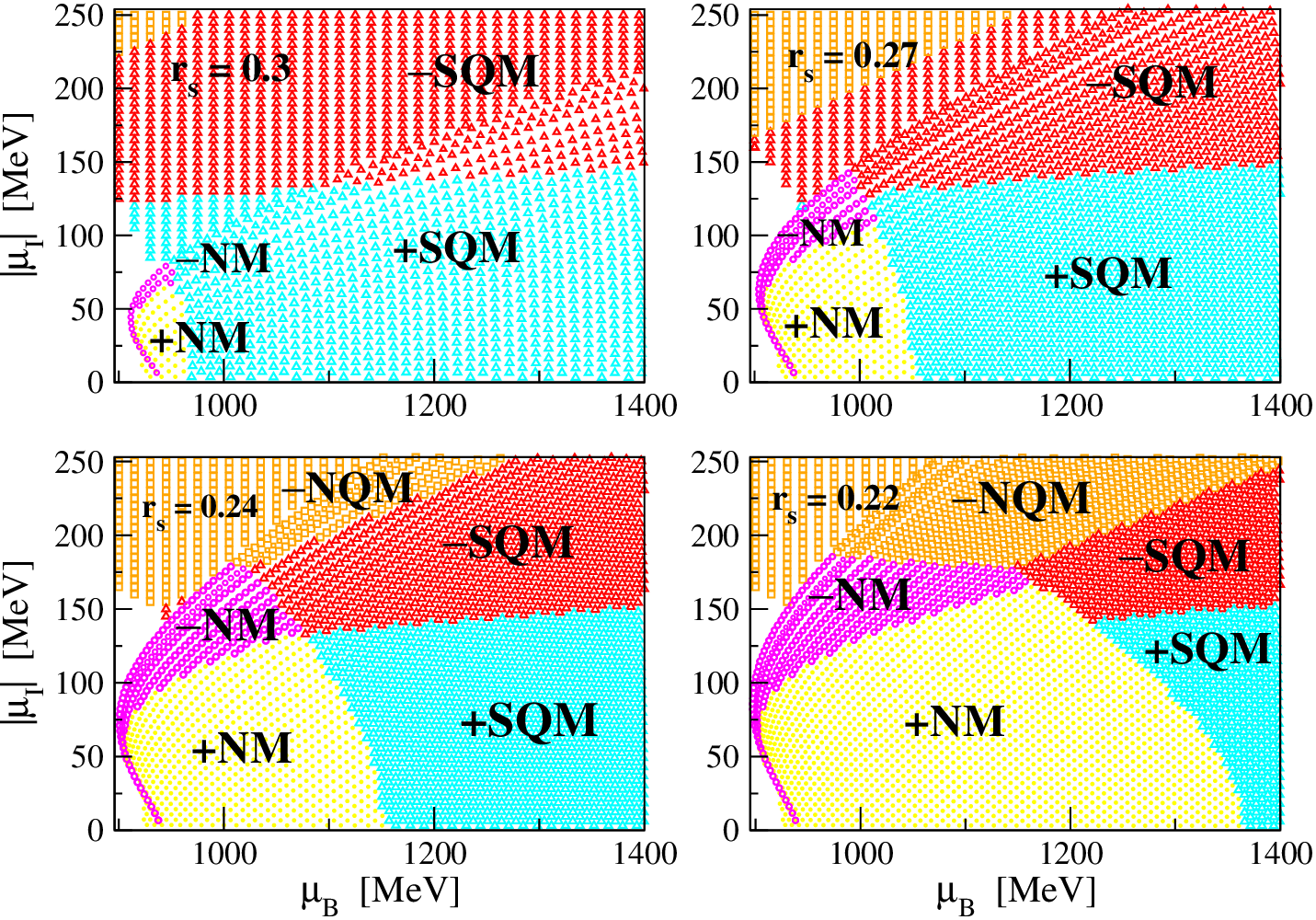}
    \vspace{8mm}  
    \caption{ Phase Diagrams with $r_{s}(NM) = r_{s}(QM)$.  The nucleon
self-energies ($\Sigma_{N}$) are 0, -200, -300 and -400 MeV respectively.
Note that in each region the sign of the charge density is indicated.  Globally charge neutral matter occurs at the boundary of the positively and negatively charged regions.}
    \label{PhaseDiagrams}
  \end{center}
 \end{figure}

\section{Phase diagrams and mixed phases}
                                                                                
To construct a phase diagram in the plane of the chemical potentials,
as in Fig.3 which is discussed below, we compare the effective potentials
for the NM, NQM
and SQM phases at each point.  The phase with the smallest
effective potential (largest pressure) is the one which is physically
realized. 
At each point we calculate the baryon and charge densities,
\begin{eqnarray}
\rho_{B}^{a} & =& -\frac{\partial V^{a}}{\partial \mu_{B}} \\
\rho_{c}^{a} & = &
 - \frac{1}{2}
( \frac{\partial V^{a}}{\partial \mu_{B}} + \frac{\partial
V^{a}}{\partial \mu_{I}}),
\end{eqnarray}
where $a =$ NM, NQM or SQM.                          
The actual equation of state as a function of baryon
density is then constructed according to global charge neutrality.
If charge neutrality can be realized within one phase (for example
the NM phase), one simply moves along the charge neutral line in the
phase diagram (for example the line +NM/-NM in Fig.\ref{PhaseDiagrams}.) as the baryon density
increases. When a phase transition occurs, it is necessary to
construct a
mixed phase, which is composed of positively and negatively charged
components belonging to two different phases \cite{Glen92}.
A charge neutral mixture of NM and QM (where QM refers to either NQM or SQM), 
for example, is characterized by the volume fraction,
\begin{equation}
\chi^{NM} = \frac{ \rho_{c}^{QM}}{\rho_{c}^{QM}-\rho_{c}^{NM}}
\end{equation}
which ranges from 0 to 1, as the density increases from the point of
pure NM
(where $\rho_{c}^{NM} = 0$) to the one of pure QM (where $\rho_{c}^{QM} = 0$).

The baryon and energy densities for the mixed phase are then expressed 
by the volume fraction as follows,
\begin{eqnarray}
\rho_{B}^{M} & = & \chi^{NM}\rho_{B}^{NM} + (1-\chi^{NM})\rho_{B}^{QM}
\label{mix1}
\\
{\cal E}^{M} & = & \chi^{NM}{\cal E}^{NM} + (1-\chi^{NM}){\cal E}^{QM}
\label{mix2}
\end{eqnarray}
Note that the components of the mixed phase have equal pressures
($P^{M}=P^{NM}=P^{QM}$) at each point ($\mu_{B},\mu_{I}$) on the
phase boundary. In this way one moves along the phase boundaries
between NM and QM while $\rho_B$ is increasing, until one comes to
the point where charge neutral pure QM is realized ($\chi^{NM}=0$)

In practice, our procedure is as follows.  
We first find the point where the effective potential for
charge neutral NM becomes equal to the one for QM.  At this point
$\chi^{NM}=1$, since $\rho_{c}^{NM}=0$.
From this point we incrementally increase either the neutron or proton
density (depending on whether the transition is in the direction of
increasing neutron and/or proton density).
For example, the transition from NM to SQM ($r_s=0.27$) is in the
direction of increasing neutron density.
For each neutron density, we determine the value for proton density
required to ensure that we are on the phase boundary.
Next we calculate the charge densities of each phase and the resulting
volume fraction.  The volume averaged properties of the mixed phase are
then related to its component phases by Eqns. (\ref{mix1}) and
(\ref{mix2}).
This process continues until we encounter the point where the QM phase
becomes charge neutral ($\rho_{c}^{QM}=0$), and thus the volume fraction
$\chi^{NM}$ goes to 0.
From here the remainder of the equation of state will be the pure charge
neutral QM phase.

\section{Results}
\setcounter{equation}{0}
                                                                                
The equations of state for this model exhibit phase transitions from the
confined quark-diquark states employed in the description of NM to a phase
with condensed quark pairs in the form of color superconducting QM.
The key point in our present work is to equate the
pairing strength for the color superconducting pairs in QM with the scalar
diquark interactions inside the nucleon.  
Depending on the amount of attractive
contributions of the pion cloud to the nucleon mass, this gives rise to a
series of phase diagrams shown in Fig.\ref{PhaseDiagrams},
corresponding to the four cases of Table 1.

\begin{figure*}
  \begin{center}
    \vspace{12mm}
    \includegraphics[width=5.7in,height=2.8in]{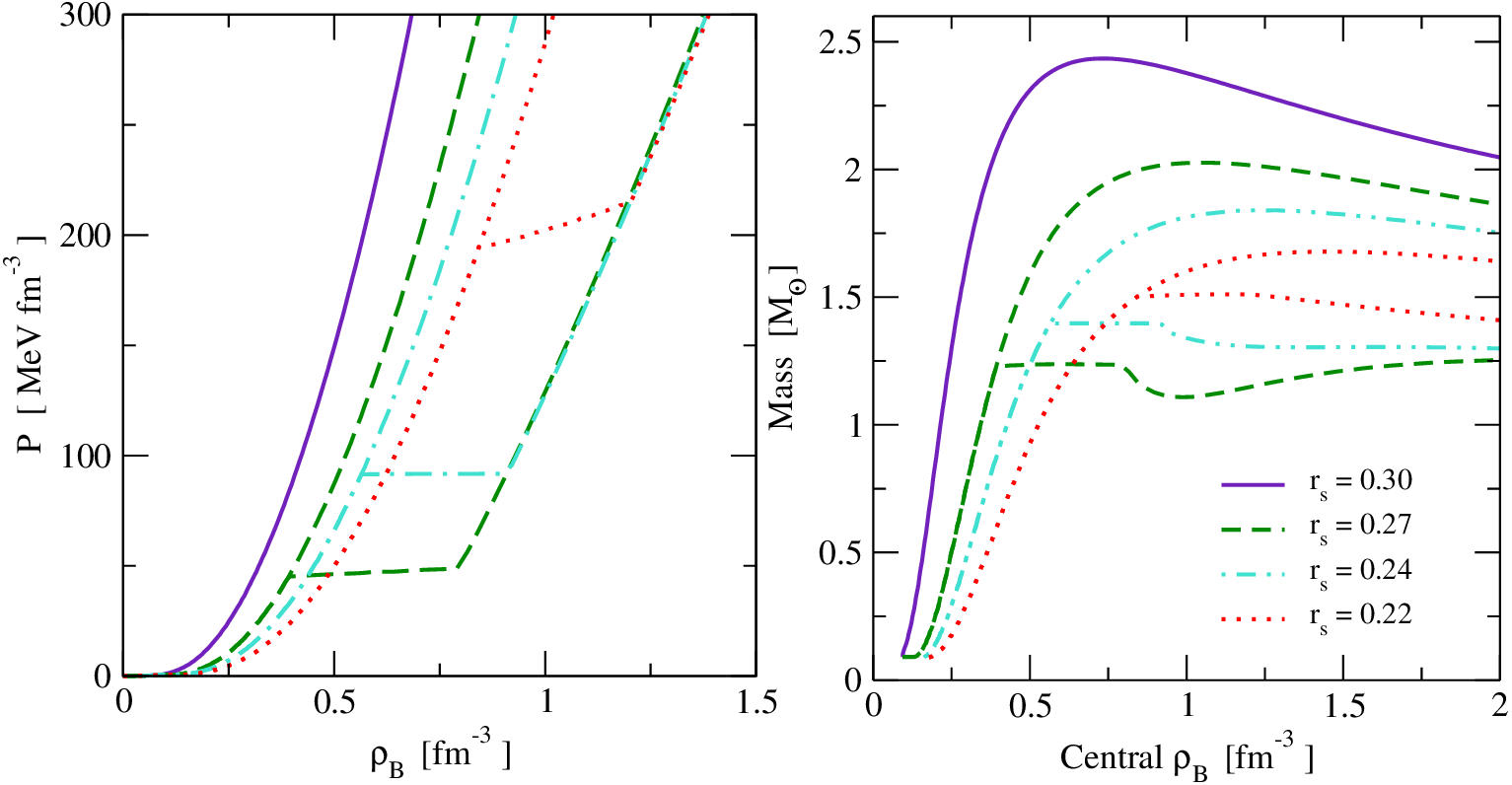}
    \caption{Equations of state and neutron star masses for the choices of scalar attraction described in the text. \vspace{4.8mm} }
    \label{stars}
  \end{center}
\end{figure*}

The four diagrams in Fig.\ref{PhaseDiagrams} are shown for increasing
pion cloud contributions to the nucleon mass, i.e., decreasing scalar
diquark interactions characterized by the ratio $r_s$.  
Starting from the first diagram in Fig.\ref{PhaseDiagrams}, the regions
occupied by SQM become smaller while those of NM and NQM phases become
larger.
In the first phase diagram, where the pion cloud is not included and the
whole attraction within the nucleon is attributed to the diquark
correlations, the NM phase is expelled almost completely and overwhelmed
by the SQM phase even in the low density region. Since this is clearly
unphysical, we can conclude that the naive quark-diquark picture of
the nucleon leads to a phase structure which is in conflict with empirical evidence.  Hence some attraction within the nucleon must be attributed to the
pion cloud.
                                                                                
The second diagram in Fig.3, which corresponds to a mass shift of
$\Sigma_N=-200$ MeV from the pion cloud, involves a mixed phase -NM/+SQM
before the system undergoes a transition to the pure SQM phase.  
The phase boundary
in this case is rather short, indicating that the pressure is almost
constant during the phase transition. In the third diagram ($\Sigma_N=-300$
MeV), the NM region extends to larger values of $\mu_B$, and the
charge neutral phase boundary shrinks almost to a point. The fourth diagram
($\Sigma_N=-400$ MeV) involves a mixed phase +NM/-SQM before the
pure SQM phase is reached.
                                                                                
The charge neutral equations of state for these four cases are shown
on the left hand side of Fig.\ref{stars}.  In the low density region ($0 > \rho_{B} > 0.001 $ fm$^{-3}$) we use the equation of state of Negele and Vautherin \cite{CRUST} to describe the neutron star crust.  
On the NM side of Fig.\ref{stars}, the equation of state
becomes softer with increasing pion cloud contributions, $\Sigma_N$.
This can be understood from our discussions in Sect. 2.  In particular, the
inclusion of the pion cloud leads to a reduction of the effective
$\sigma NN$ coupling in the medium, which must be balanced by
a smaller vector coupling (the parameter $r_{\omega}$ in Table 1)
in order to maintain the correct binding energy at the
nuclear matter density $0.17$ fm$^{-3}$.  Note that the actual saturation point for the model moves to somewhat higher densities (0.14, 0.19, 0.21 and 0.25 fm$^{-3}$) 
when $\Sigma_{N}$ increases (0, -200,-300 and -400 MeV respectively).  As noted in earlier works the only parameter in the present model that may be adjusted to give the saturation point is $G_{\omega}$ \cite{Ishii95}.  
With this in mind these results are reasonably close to the empirical value of $\rho_{0} = 0.17 $ fm$^{-3}$. 

On the QM side all cases considered here give almost identical equations
of state, indicating that the relationship between density and pressure in charge neutral SQM is not sensitive to the value of $r_s$ (or indeed to the value of $\Delta$).  
However, the transition densities are sensitive to $r_s$. 
In all cases considered here, 
the pressure variations in the mixed phase are rather small, 
indicating that our construction, based on the Gibbs criteria of phase equilibrium, gives similar results to a naive Maxwell construction between
charge neutral NM and SQM.

Through the Tolman Oppenheimer Volkoff (TOV) equations \cite{TOV} any equation
of state specifies a unique set of non-rotating relativistic stars.
The right hand side of Fig.\ref{stars} illustrates the solutions to
these equations for each of the equations of state on the left hand side
of Fig.\ref{stars}.
We see from these figures that the inclusion of pion cloud contributions to
the nucleon mass, leading to a decreasing strength of scalar diquark pairing,
has significant effects on the properties of neutron stars.
Because of the softening of the equation of state, the maximum masses for
pure hadronic stars are reduced from 2.4 solar
masses to 2.0, 1.8 and 1.7 solar masses as $\Sigma_{N}$ increases (and
$r_{s}$ decreases accordingly).  

\begin{figure*}
  \begin{center}
    \vspace{8mm}
    \includegraphics[width=2.8in,height=2.6in]{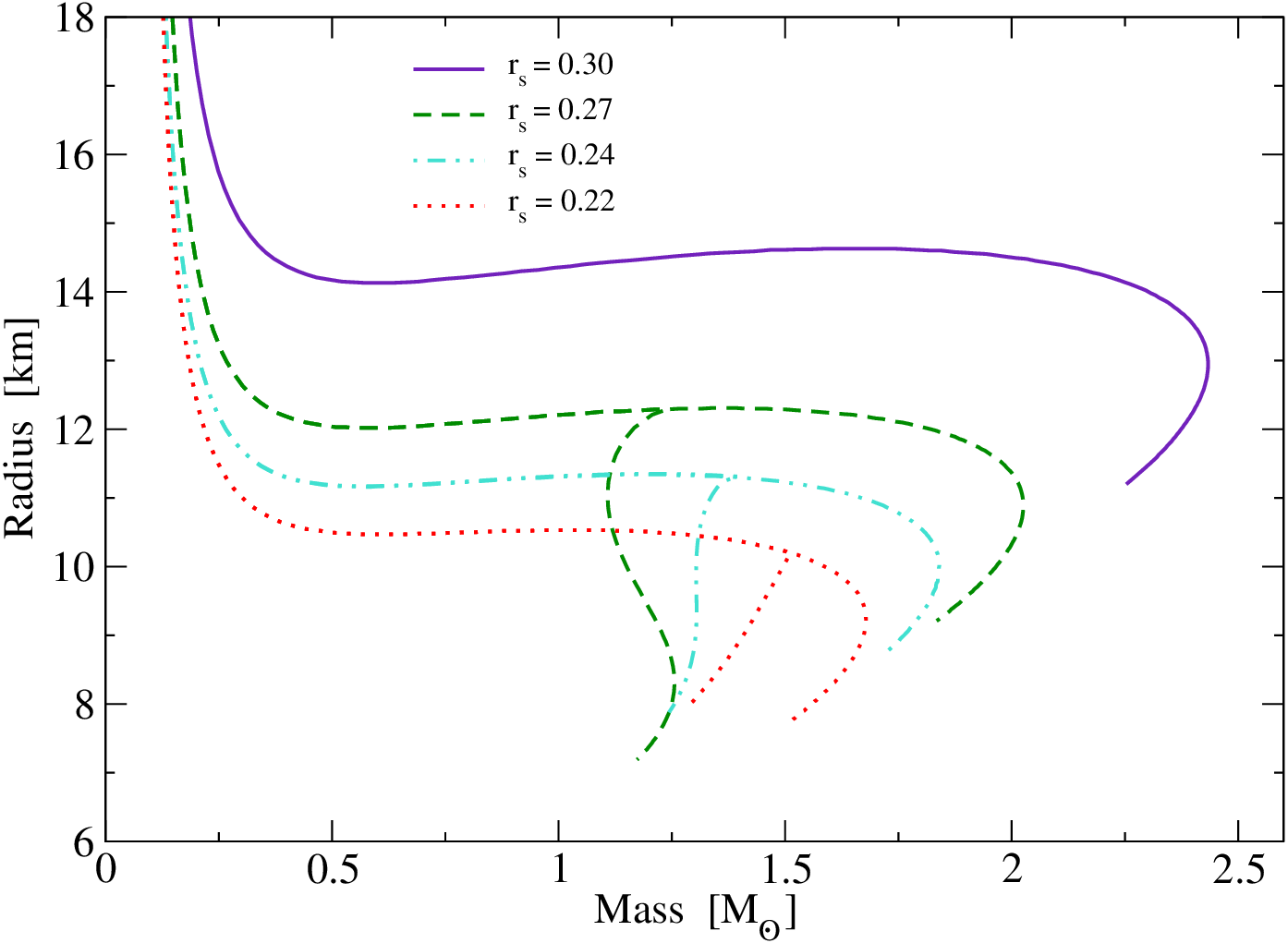}
    \caption{Mass radius relationship of neutron stars and hybrid stars for the choices of scalar attraction described in the text.}
    \label{MassRad}
  \end{center}
\end{figure*}

The equations of state with phase transitions to SQM produce plateaus in the central density vs mass curves in Fig.\ref{stars}.  
In the case of $r_{s} = 0.22$ only the mixed phase can be reached inside a star, since a region of negative slope in this plot corresponds to unstable solutions to the TOV equation.  However, stable hybrid stars with quark cores are possible in the case of $r_{s} =
0.27$ and $r_{s} = 0.24$.  
In these configurations there are additional sets of stable solutions at higher densities, which in the literature are refered to as twins  \cite{Glen98}(since for these configurations there can be stars that have the same mass but different radii, as illustrated in Fig.\ref{MassRad}).  Qualitatively similar results are also found in Ref.\cite{Fraga01}.
It has recently be shown that the second set of solutions are indeed stable and may give rise to an observable signature for the occurance of phase transitions in compact stars \cite{Schertler00}.

However, in the case of $r_{s}=0.27$ the masses are too small to allow for
observed pulsar masses, which are typically about 1.4 solar masses.  
For $r_{s} = 0.24$ and $r_{s} = 0.22$ the maximum masses are 
approximately 1.4 and 1.5 solar masses, respectively.  
One possibility that may give rise to more massive hybrid stars in
this model is to delay the phase transition to QM.  If hyperons were
included, for example, then the
transition may be shifted to higher densities, since the equation of state
in the hadronic
phase would be softened independent of the value of the pairing strength
$r_{s}$.
We note that the correlation between the transition densities and the maximum neutron star masses shown in Fig.\ref{stars} follows the phenomenological discussions on phase transitions given in Ref.\cite{Takahara}.

It is interesting to note that these phase transitions to QM give rise to
plateaus in the neutron star masses.  This phenomenon may be the
reason that so many observed neutron stars lie within such a narrow mass
range \cite{Lattimer}.
In addition, Fig. \ref{MassRad} shows that the radii of the stars may be reduced by 2 - 4 km if the pion cloud contribution to the nucleon mass
increases (and the scalar pairing interaction decreases accordingly).

\section{Discussion}
\setcounter{equation}{0}
                                                                                
We have used the NJL model, supplemented with a method of regularization which simulates confinement, to calculate consistently the properties of
nuclear matter and quark matter.  In doing so we have related the quark
interactions in the confined phase, to the quark interactions in the
deconfined phase, where color superconductivity is assumed to arise.
Our principal finding is that the scalar pairing between quarks within
the nucleon and in QM may be equated, if the attraction within the
nucleon is attributed not only to the diquark interactions but also
to the pion cloud. Since the attraction in the scalar $\overline{3}_c$
channel between (almost) current quarks at high
densities or energies can be derived directly from QCD,
this result lends some support to the Faddeev approach to the nucleon,
since the quark-diquark picture may be characterized by not only the same
type of pairing interaction but also the same strength as we expect to
find in the high density QM phase.
This is an important feature of nucleon dynamics and is relevant to any
finite density studies that incorporate nucleon structure.
                                                                                
By including the pion cloud contributions to the nucleon mass, we found
that the equation of state of nuclear matter becomes softer, reducing the
neutron star and hybrid star masses significantly.

\vspace{1pc}
                                                                                
{\sc Acknowledgments}
                                                                                S.L. would like to thank Ian Cloet for helpful discussions.
This work was supported by the Australian Research Council and DOE
contract DE-AC05-84ER40150,
under which SURA operates Jefferson Lab, and by the Grant in Aid for
Scientific Research of the Japanese Ministry of Education,
Culture, Sports, Science and Technology, Project No. C2-16540267.

\end{document}